%% file: ms.tex
\documentclass[preprint]{aastex631}

\tighten

\shorttitle{WiFeS Observations of Puppis A}
\shortauthors{Ghavamian et al.}
\usepackage{graphics,graphicx}
\usepackage{todonotes}
\usepackage{comment}

\def\kms{km\,s$^{-1}$}

\def\wifes{WiFeS\,}

\begin{document}

\title{The Peculiar Ejecta Rings in the O-Rich Supernova Remnant Puppis A: Evidence of a Binary Interaction?}

\correspondingauthor{Parviz Ghavamian}
\email{pghavamian@towson.edu}

\author[0000-0002-9886-0839]{Parviz Ghavamian}
\affiliation{Department of Physics, Astronomy and Geosciences,
  Towson University, Towson, MD, 21252; pghavamian@towson.edu}
  
\author[0000-0002-5044-2988]{Ivo R. Seitenzahl}
\affiliation{School of Science, University of New South Wales,
Northcott Drive, Canberra, ACT, 2600, Australia; 
i.seitenzahl@adfa.edu.au}

\author[0000-0002-9665-2788]{Fr\'ed\'eric P.~A.~Vogt}
\affiliation{Federal Office of Meteorology and Climatology MeteoSwiss, Chemin de l'Aerologie 1, 1530 Payerne, Switzerland; frederic.vogt@alumni.anu.edu.au}

\author[0000-0002-7868-1622]{John C. Raymond}
\affiliation{Harvard-Smithsonian Center for Astrophysics, 60 Garden Street, Cambridge, MA, 02138, USA;
raymond@cfa.harvard.edu}

\author{Jasper Scelsi}
\affiliation{Department of Physics, Applied Physics and Astronomy,
  Jonsson Rowland Science Center, Troy NY, 12180; scelsj@rpi.edu}

\begin{abstract}
Near the center of the Puppis A SNR a series of nested optically emitting rings of high velocity ejecta (known as `the Swirl') were identified several decades ago by \cite{1989Natur.337...48W}. To date no follow-up observations of these rings have been published, and their physical origin has remained a mystery.   We present results of integral field spectroscopy of the Swirl using the Wide Field Integral Spectrograph (\wifes) on the 2.3\,m telescope at Siding Spring Observatory in Australia.    The outermost ring exhibits a nitrogen-rich spectrum blueshifted to 1350 km/s, with smaller blueshifted rings within the first exhibiting mostly oxygen-rich spectra moving at 1000 km/s and 750 km/s.  The structures are connected by material of intermediate velocity and variable composition, including sulfur-rich material.  The Swirl is turbulent and shock excited, and contains as much as 0.5 M$_{\odot}$ of metal-rich material.  The chemical composition and exclusively blueshifted radial velocities of the Swirl are consistent with progressively deeper nucleosynthetic layers in a massive progenitor star.  We suggest the possibility that the Swirl marks a `funnel' carved into the supernova ejecta by a close, massive binary companion at the moment of explosion.   

\end{abstract}


\section{Introduction}

Oxygen-rich supernova remnants (OSNRs) are remnants exhibiting prominent oxygen forbidden-line emission at optical wavelengths.  They were first discovered with the optical studies of Cassiopeia A by \cite{1971ApJ...167..223P} and \cite{1978ApJ...219..931C}.  The emission is often distributed in filaments and knots, and is understood to arise from shocked ejecta of a core-collapse supernova (SN). Unlike most SNRs, where the optical emission comes from interstellar medium (ISM) ionized by the forward shock, O-rich SNRs are special in that one can study the actual ejecta composition of the supernova, providing unique and direct insights into explosion mechanisms and nucleosynthesis conditions in core-collapse supernovae. Only a handful of such O-rich SNRs are known, including Puppis A, Cassiopeia A and G292.0+1.8 in our Galaxy, N132D and 0540$-$69.3 in the LMC, 1E0102$-$7219 in the SMC and a luminous supernova remnant in NGC 4449  \citep[see][and references therein.]{blair00} 

\begin{figure}[ht] 
 \centering
\includegraphics[width=6.5in]{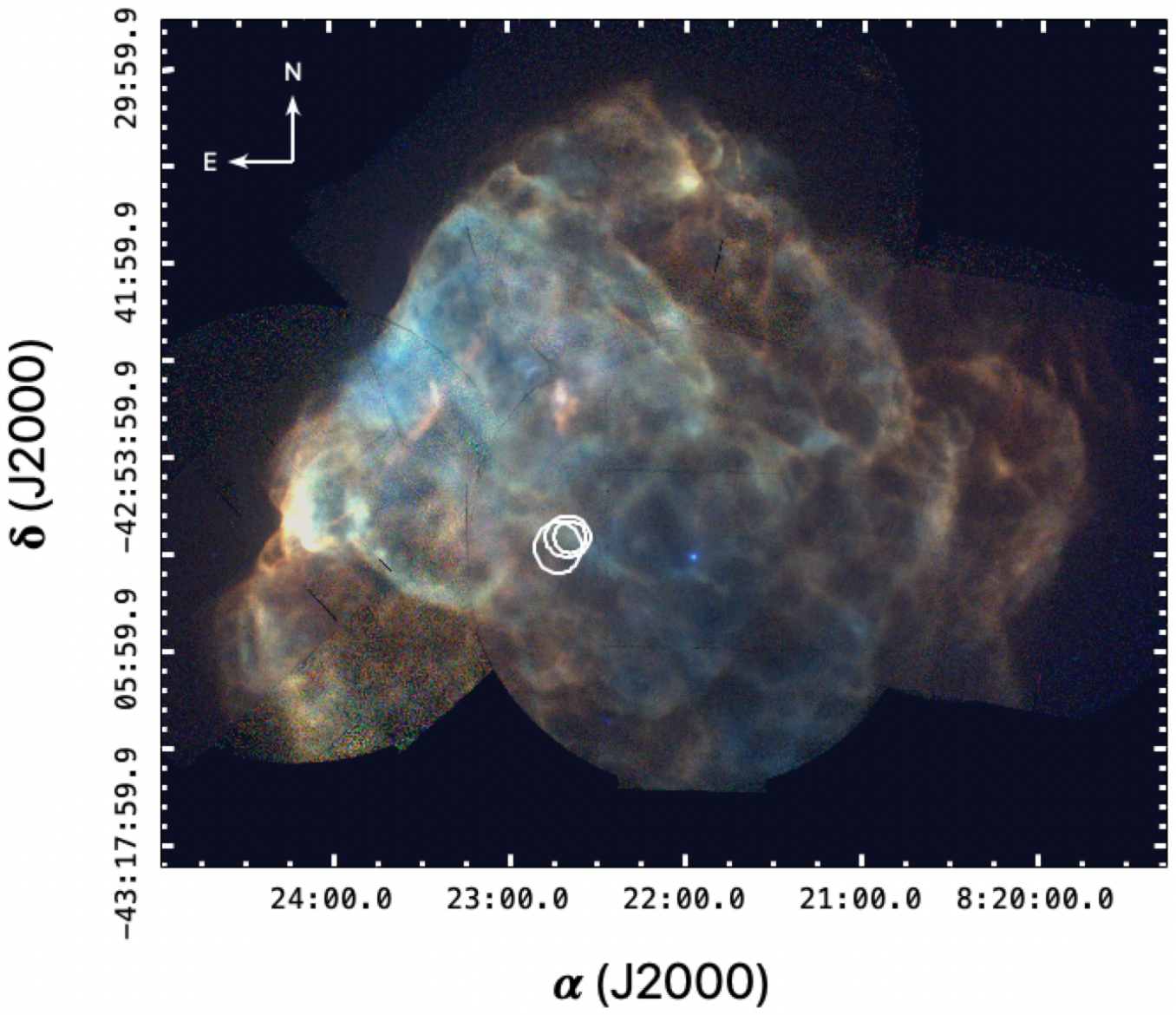}  
   \caption{ The XMM image mosaic of Puppis A (Katsuda et al. 2010) shown with the location and approximate shape of the Swirl marked (white ellipses).  The colors indicate the following XMM energy bands: 0.5-0.7 keV (Red), 0.7-1.2 keV (Green) and 1.2-5.0 keV (Blue).  The CCO J0822$-$4300 appears as the bluish point source to the west of the Swirl. }
   \label{fig:fig1}
\end{figure}

Puppis A is one of the brightest SNRs in X-rays, with a large angular diameter of around 50${\arcmin}$.  The presence of a central compact object (CCO) \citep{1996ApJ...465L..43P} in Puppis A, as well as abundance analyses of its ejecta, indicated that it was the product of a core-collapse SN.  Although its X-ray appearance is dominated by its interaction with ISM clouds (especially on the eastern and northern rims, e.g., \citealt{1987ApJ...314..673T}, \citealt{2005ApJ...635..355H}) (Figure \ref{fig:fig1}), several structures with enhanced abundances of O, Ne and Mg were identified by \cite{Katsuda2008}, who suggested that this material was SN ejecta from the massive progenitor.   However, prior to this discovery, evidence of SN ejecta in Puppis A had already been uncovered by \cite{1985ApJ...299..981W}, who found fast-moving ejecta emitting mostly in [O~III].  Along with this material (named the $\Omega$ filament due to its morphology, \citealt{1985ApJ...299..981W}),  \cite{1988srim.conf...65W} discovered more optically emitting ejecta near the $\Omega$ filament, and inferred a dynamical age of 3700$\pm$300 years for Puppis A based on the proper motions of the oxygen-rich material.    
The presence of optical emission from O-rich material in Puppis A was a surprise, since at an advanced age of 3700+ years, the denser ejecta knots should have all dissipated by now.  

Puppis A also shows prominent filaments of slow moving radiatively shocked material with enhanced abundances of nitrogen and sulfur.  These filaments were first uncovered in the optical surveys of Puppis A \citep{1954ApJ...119..206B}.  In followup imagery and spectroscopy \citet{1977ApJ...214..179D} suggested that this material was left over from the AGB phase of the progenitor star, and that it coagulated into clumps when a fast Wolf-Rayet wind swept up the material. 

Multiwavelength studies have shown that the Puppis A SN explosion was asymmetric, with the CCO (RX J0822$-$4300) receiving a high velocity kick toward the SW (see \citealt{mayer20}) and the O-rich ejecta (as traced by the [O~III] emission, see \citep{1985ApJ...299..981W} and \citep{1988srim.conf...65W}) being expelled in the opposite direction, toward the NE.  

The origin of one particularly interesting feature in Puppis A located approximately 7$^{\prime}$ south of the $\Omega$ filament and discovered in narrowband optical images by \citet{1989Natur.337...48W} has remained a mystery. This is the group of nested ring-shaped structures located near the geometric center, known as `the Swirl' \citep{1989Natur.337...48W}.   Based on their narrowband [O~III], $H\alpha$+[N~II] and [S~II] images, as well as followup longslit spectroscopy, \cite{1989Natur.337...48W} found that the rings exhibited a high blueshifted radial velocity ($|v|$ $\geq$ 1000 km/s) as well as enhanced abundances of N, O and S.  They hypothesized that the Swirl may be the remnant of a second supernova that exploded within the last 1000 years inside of Puppis A.  

In this paper, we present new integral field spectra of the Swirl, which allow us to dissect the kinematics and trace the abundance variations of the ring material in unprecedented detail.  While confirming the earlier findings of \cite{1989Natur.337...48W} of metal-rich composition of the Swirl, we also show that the kinematics are inconsistent with limb brightened spherical shell expansion, and instead indicate that they are expanding in a conical geometry consistent with a tunnel carved out of the expanding supernova ejecta.

\section{Observations and Data Reduction}

\wifes is an integral field unit (IFU) located at the Nasmyth A focus of the Australian National University 2.3m telescope located at Siding Spring Observatory, Australia.  The IFU data were acquired in `binned mode', providing a field-of-view of $24 \times 38$ spatial pixels (spaxels) each 1${\arcsec}\times$1${\arcsec}$ in angular size, giving a field of view of 24${\arcsec}\times$38${\arcsec}$.  The instrument is a double-beam spectrograph, with a dichroic beamsplitter providing independent channels for each of the blue and the red wavelength ranges, and data are collected simultaneously in both channels during each exposure.

 We performed the \wifes observations of the Swirl in Puppis A on 4 nights in 2017 (26 and 27 February and 1 and 2 March) (Proposal ID 1170247; PI Seitenzahl) and 6 nights in 2018 (11, 13, 14, 15, 16, 17 February) (Program ID 1180105; PI Seitenzahl). A total of 29 fields were observed.  Most fields were co-added in two 1800\,s  exposures, for a total integration time on source per field of 1 hour.  In other fields there was a reduced on-source integration time of 900 s per exposure.  In these cases, additional exposures were taken to bring the total integration time for those fields to 1 hour.  Immediately before and/or after the science exposures we observed dark patch of sky for 2$\times$900\,s.  We used these ``sky" exposures to subtract the night sky emission during the data reduction step. Every night, we also observed the standard star LTT4364 for 2$\times$300s for flux calibration, as well as a daily set of bias, flat-field, and arc-lamp (Cu-Ar) calibrations.
 
 In all observations we used the B3000 gratings for both channels, covering together (and simultaneously) the 3500 -- 7000 \AA\, wavelength range.  These provided a resolution of R = 3000 in the blue (3500 -- 5700 \AA) and R = 3000 in the red (5300 -- 9560 \AA) \citep{2007Ap&SS.310..255D}.  The corresponding velocity resolutions are 110 \kms\ in both channels.

Since the angular size of the Swirl exceeds the \wifes\ field-of-view, we mapped it using 29  overlapping fields (Figure~2).    After sky subtraction we separately corrected for the OH and H$_2$O telluric absorption features in each frame, then used the spectrophotometric standard LTT4364 \citep{1999PASP..111.1426B} to perform the absolute photometric calibration of the data cubes. Transparency conditions varied by less than 10\% during the Swirl observations and the seeing varied between 1\farcs5 and 2\farcs5.  We regularly acquired internal wavelength calibration and bias frames during the observations, and obtained internal continuum lamp flat fields and twilight sky flats in order to provide sensitivity corrections in both the spectral and spatial directions. Due to the faintness of the flux calibration standard in the blue, the flux calibration of the blue channel is less accurate than for the red channel, being least accurate below 4000 \AA.  

We reduced the data using \textsc{pywifes} v0.7.3 reduction pipeline \citep{2014Ap&SS.349..617C}. The pipeline produced a wavelength calibrated, sensitivity corrected, photometrically calibrated data cube corrected for telluric features and cosmic ray events. The \textsc{pywifes} pipeline uses an optical model of the spectrograph to provide the wavelength calibration for each channel, resulting in a wavelength solution valid across the entire detector.  In doing so, each pixel on the CCD is assigned a precise wavelength and spatial coordinate (i.e., a spaxel) by the data reduction pipeline.  The final data cubes (one per exposure and per spectrograph channel) are then reconstructed and interpolated onto a regular three-dimensional grid with wavelength dispersion of 0.768 \AA\, pixel$^{-1}$ in the blue and 0.439 \AA\, pixel$^{-1}$ in the red. 

The respective alignment of the different \wifes fields in the mosaic was derived by comparing the reconstructed summed continuum frames from the red cubes with the Digitized Sky Survey 2 (DSS 2) red band image of the area. Given the mean seeing conditions during the observations ($\sim$2\farcs0), the spatial shifts between each individual cube are rounded to the nearest integer for simplicity, and also to avoid superfluous resampling of the data. All data cubes are forced onto the same wavelength grid in the data reduction process, so that no shift is required along that axis. The final mosaic (including gaps) has dimensions 3\farcm6$\times$4\farcm 1 on the sky, which for an assumed distance of 2.2 kpc for Puppis A \citep{2003MNRAS.345..671R} corresponds to a physical size of 2.3$\times$2.9 pc (though a more recent assessment by \cite{2017MNRAS.464.3029R} from H~I observations finds 1.3$\pm$0.3 kpc). The WCS information in the FITS headers was pegged to the coordinates of a star from the 2MASS catalogue present in the mosaic field, resulting in an overall positioning of the mosaic and the respective alignment of each field accurate to $\lesssim$1\arcsec. 


\section{Broad Characterics of the Puppis A Rings}

The Swirl was discovered and first described by \citet{1989Natur.337...48W} in narrow band optical images of Puppis A.  The structure is located approximately 3\farcm7 SE of the expansion center determined by \citep{1988iue..prop.3202W}.   It appears as a set of nested, partially complete and overlapping shells of differing radii, with the largest nearly 3\arcmin\  across.   The rings are offset approximately 8\farcm5 E of the well known CCO J0822$-$4300 (Becker et al.\, 2012).  In Figure~1 the XMM mosaic of Puppis A (Katsuda et al. 2010) is shown with the approximate outlines of the Swirl marked.  There is no discernible X-ray counterpart to the Swirl in the XMM images, nor are there any X-ray emitting ejecta knots (as identified in prior work by \citet{2010ApJ...714.1725K} and \citealt{Katsuda2008}) seen in close proximity to the Swirl.  This is in contrast to structures such as the O-rich $\Omega$ filament \citep{1985ApJ...299..981W}, which has no direct X-ray counterpart but is connected to an extended O-rich structure of similar chemical composition in the X-rays \citep{Katsuda2008}.  

\begin{figure}[ht] 
 \centering
\includegraphics[width=5in]{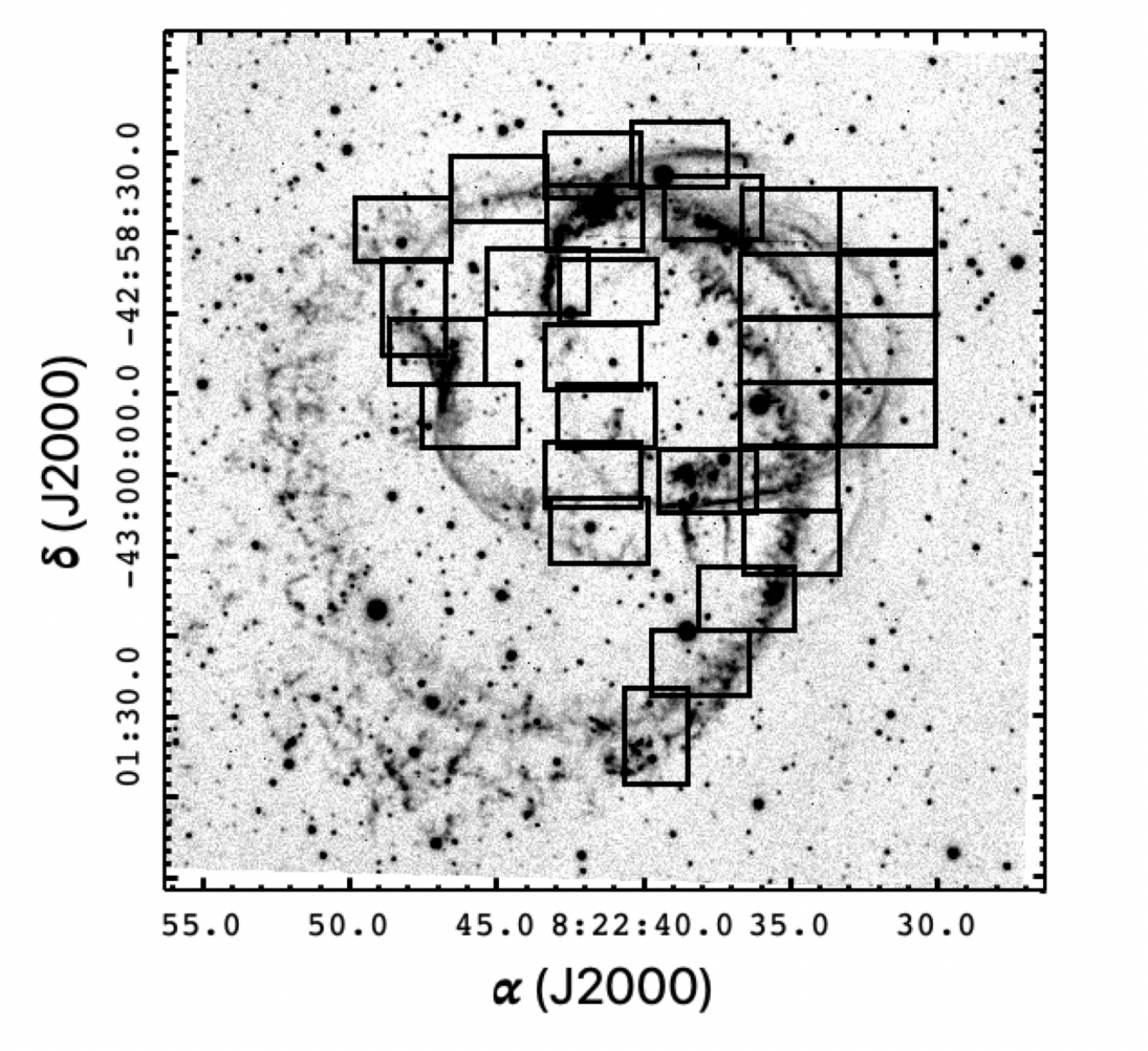}  
   \caption{An H$\alpha$+[N~II] image of the Swirl in Puppis A.  The image was acquired in 2010 with the 4m Mosaic imager of CTIO (NOIRLab archival image (PID 2010A-0483)).  The 29 fields covered by the WiFeS observations are marked.  Each field is 24$^{\prime\prime}\times$36$^{\prime\prime}$.  North is up and East to the left.   }
   \label{fig:fig2}
\end{figure}

A closeup view of the Swirl in H$\alpha$+[N~II] is shown in Figure \ref{fig:fig2}.    The image shows that the largest ring is slightly squashed in the E-W direction (3\arcmin $\times$ 3\farcm2 across), giving the impression it is slightly inclined to the line of sight, while the two smaller rings (2\farcm8 $\times$ 2\arcmin \, and 1\arcmin $\times$ 1\arcmin) are broken and distorted, and centered at a point offset from the largest one.    The eastern half of the largest ring, which appears highly fragmented, was not covered by our \wifes observations.  This was due to the faintness of the fragments, the large area covered by them, and constraints on the observing time.    The \wifes detector footprints for the individual fields are marked on Figure~\ref{fig:fig2}.

\section{Major Emission Lines}

A cursory inspection of the \wifes\, data shows that the Swirl exhibits a wide variety of low and high ionization nebular emission lines from several elements (H, He, N, O, S and Fe and Ni), as first reported by \citet{1989Natur.337...48W}.    Furthermore, each line is detected at a wide range of highly Doppler shifted velocities.  All of these properties make the detailed characterization of the Swirl very challenging. 

To visualize and spatially map the emission from each ring we utilized QFitsView \citep{2012ascl.soft10019O}, a software package designed for visualization and spectral extraction of integral field data. The high quality spectra from each spaxel of the WiFeS datacube allow us to generate continuum-subtracted images in prominent emission lines at all velocities, effectively acting as a tunable narrow band pass filter.  After careful visual inspection of hundreds of spectra in QFitsView, we created images in each emission line by first summing the datacube flux over a wavelength interval covering the line.  We removed the underlying continuum in QFitsView by averaging the continuum over wavelength windows on either side of each emission line, then subtracting the averaged flux of the two continuum windows from the emission line window.  This produced images of the rings in each emission line with most stars removed and the sky background near zero (though it did not remove background emission lines when they overlapped with object emission lines $-$ as we describe later we performed this task separately for lines affected by this overlap).

Narrowband images of the Swirl in each detected emission line are shown in Figure~3.  We found that the outermost ring emits strongly in [N~II], with all other lines from that ring at least an order of magnitude fainter.  The interior (smaller) rings, however, while showing significant [N~II], also exhibit strong emission from other lines.  Though considerably fainter in some locations than others, the inner ring is discernible in all of the forbidden oxygen lines ([O~I], [O~II], [O~III]) as well as [N~II], [S~II] and H$\alpha$.  Emission from He~I $\lambda$5876, [Ca~II] $\lambda$7291 and [Fe~II] $\lambda$8616 are also detected in an isolated segment of the northern part of the Swirl.  Interestingly, this isolated segment (most clearly seen in the [N~II] image) overlaps with the outermost ring and then extends eastward into another bent ring-shaped structure.  This structure was only partially covered in our observations, though the missing part of that ring can seen in Figure~2, where it curves southward and then splits into multiple strands as it fades away.  Although all three rings are also visible in H$\alpha$, very little corresponding H$\beta$ emission is seen overall save for the brightest segments of the inner ring (consistent with the significant extinction found toward Puppis A in earlier studies).

Seen from Earth, there is considerable line of sight overlap between the rings of the Swirl, making it difficult to discern whether they are physically connected.  However, both their radial velocities and their composition (as revealed by Doppler shifts and emission line ratios) follow a coherent pattern which allows for some disentanglement of their spatial components.  In Figure \ref{fig4} radial velocity maps of the Swirl are shown in the light of the three most prominent emission lines: [N~II] $\lambda$6548, [O~III] $\lambda$5007 and [S~II] $\lambda$6716.  


\section{Spectral Extraction and Line Fitting}

The WiFeS datacube of the Swirl shows a dynamic, highly clumpy structure with a complex morphology.  A close examination of the rings shows them to be clumpy on scales of at least two spaxels (approximately 2${\arcsec}$ square). To capture the full range of spectral variability present in the rings, we extracted red and blue channel spectra from a large array of adjacent boxes 9 square spaxels in size (3${\arcsec}\times\,$3${\arcsec}$), large enough to account for the atmospheric seeing but small enough to sample variations along the rings.

All spectra used in deriving the velocity maps in Figure~4 were extracted using self-written python routines, which for each specified extraction (x,y) spaxel computed the average flux per spaxel within the 3$\times$3 spaxel extraction box as well as the corresponding average flux variance per spaxel.  Emission lines were fit using self-written python routines and a python implementation of the MPFIT algorithm \citep{2009ASPC..411..251M}.  Gaussian profiles and a linear background were used having variable height, width and centroid.  For [N~II] approximately 725 separate spectra were fit with Gaussians using chi-squared minimization, with the weaker of the two [N~II] lines chosen in order to avoid confusion between [N~II] $\lambda$6583 and background H$\alpha$, which overlapped at a number of locations due to blueshifting of the former line in the supernova ejecta.  This line was also used because no other emission lines were present up to 20 \AA\, blueward of the line at 6548 \AA, which allowed for useful estimate of the baseline for Gaussian fitting.  For spectra exhibiting one [N~II] $\lambda$6548 component, a single component Gaussian was fit along with a local baseline.  

We performed line profile fitting in a series of runs, with each run centered on the wavelength interval where the targeted line profile was expected to be found.  A cursory examination of the ring spectra showed that all of the emission lines were substantially blueshifted and typically clustered around three approximate radial velocities: $\Delta v$ = $-$1350 \kms, $-$1025 \kms\, and $-$750 \kms.  For each emission line we set up three separate runs with wavelength intervals 20\AA\ wide and centered on the wavelength of each velocity shifted line.  We estimated initial parameters for each Gaussian in the following way.  We initially guessed the peak by finding the maximum value of the specific flux in the fitting wavelength window, while setting the initial line width to the instrumental resolution ($\lambda_{\sigma}\,=\,\frac{\lambda_0}{3000}$).  We guessed the initial wavelength centroid via  $\lambda_i\,=\,\lambda_0\,(1\,+\,\frac{\Delta v}{c})$, where $\Delta\,v$ is one of the three radial velocities listed above.  The background level was initially set to 10$^{-17}$ ergs cm$^{-2}$ s$^{-1}$ \AA$^{-1}$ arcsec$^{-2}$.

We inspected all fitted line profiles by eye and for those giving unsatisfactory results, the initial parameter estimates were tweaked as necessary.   We then reiterated the line profile fitting until all line profile fits looked satisfactory.  Although single line profile fitting was sufficient for determining the radial velocities for the [S~II] and [O~III] lines, in approximately 10\% of the spectra the [N~II] emission consisted of two Doppler shifted components.  This mostly occurred in locations where the outer ring overlapped with an inner ring having slightly lower radial velocity, for example in the $-$1000 - $-$1100 \kms\, range (as opposed to the outer ring at $-$1350 \kms).  In those cases the line profile overlap required double Gaussian fitting.    

In a handful of cases (mostly in the outer ring, $\Delta v$ = $-$1350 \kms, the two [N~II] lines were separated by less than the FWHM of each line, suggesting that two components were needed even though the spectra were from a single ring.   A number of these line profiles exhibited a FWHM slightly exceeding the instrumental resolution (110 \kms) and exhibited weak (though noticeable) high velocity tails extending blueward of the line.  Variations in the FWHM can be seen in the velocity maps in Figure~4, where the size of each velocity point reflects the line profile width.  These characteristics are consistent with turbulence and shear motions within the Swirl, which impart additional dispersion in the radial velocities within the extraction window.  

The velocity distribution of the rings in Figure~4 shows that they are blueshifted to velocities in the range 600 \kms $\leq$ $|\Delta v|$ $\leq$ 1350 \kms.  Furthermore, the [N~II] emission, which occurs most extensively throughout the rings, shows that, allowing for some dispersion, the rings can be roughly separated into three categories: the outermost (largest) ring at $\Delta v$ $\sim$ $-$1350 \kms, a slightly smaller intermediate ring at $\Delta v$ $\sim$ $-$1000 \kms\, and innermost (smallest) ring at $\Delta v$ $\sim$ $-$750 \kms.  

The patterns in radial velocity are even more easily seen in a histogram of the radial velocities, as shown in Figure~5. There are three identifiable trends from Figures 4 and 5: (1) the radial velocities of the rings are almost linearly proportional to their radii, with larger rings having higher average radial velocity, (2) as can be judged by the presence or absence of S, N and O emission in each ring,  oxygen emission is weakest in the outermost ring and strongest in the innermost ring, and (3) rings containing the heaviest nucleosynthetic products occur at lowest radial velocity. The N-rich material (where spectra exhibit [N~II](6583)/H$\alpha$ ratios well in excess of unity) can be found in all three rings, while the O-rich and S-rich material clusters around $\Delta v$ $\sim$ $-$750 \kms.  

Example spectra from one of the rings is presented in Table~1, where the measurements were obtained from the position labeled P1 (location marked by the arrow in Figure~3 is shown).  We present measurements from this particular spectrum because it is one of the few locations showing all the major lines, both in the blue and in the red (most locations show either strong N and weak O and S or vice versa).  Here we detected weak lines of He and Fe for the first time  (though these lines have also been seen in other O-rich ejecta, e.g., \citet{1985ApJ...299..981W}). We also detect weak lines of [Ar~III] $\lambda$7135, [Ca~II] $\lambda$7291 and [Ni~II] $\lambda$7378, which to our knowledge are the first time these lines have been seen in Puppis A. 

The P1 spectrum, acquired from one a position of intermediate velocity ($\sim$-1025 km/s) exhibits a [N~II](6583)/H$\alpha$ ratio of around 8.4, indicating enhanced nitrogen abundance over solar.  Such a high value is consistent with similarly high values measured at other locations in Puppis A (\citealt{1977ApJS...33..437D, 1983IAUS..101..193D}).  For comparison, this ratio does not exceed 1.0 in Galactic SNRs in interstellar gas  (see, for example \citealt{1996ApJS..106..563F}).  The [S~II]($\lambda$6716+$\lambda$6731)/H$\alpha$ ratio here is nearly unity, which is somewhat higher than the average value in the rest of the rings.  

\section{Density Structure of the Swirl}


The reduced WiFeS spectra of the Swirl show forbidden emission lines over a wide range in ionization stages, as would be expected for shocked material. Moreover, specific emission lines show strongly enhanced emission from some species, while exhibiting little to no emission from others.  This is consistent with the finding of \citet{1989Natur.337...48W} of strong variations in metal abundances in the rings.   

In addition to the radial velocities of the rings, their densities are also relevant for establishing their physical origin.  Most importantly, when combined with abundance estimates from line ratios, their densities can be used to constrain the total mass of the rings, and hence shed light on the physical mechanism behind their formation.  The [S~II] $\lambda$6716/$\lambda$6731 density-sensitive ratio is a well known nebular diagnostic, and fortunately  [S~II] is detected from most of the rings (c.f. Figure~\ref{fig3}).  To measure this density-sensitive ratio throughout the rings we produced `narrowband' [S~II] images separately for each line of the doublet, and separately for each of the three main radial velocities ($-$1350 \kms, $-$1025 \kms\, and $-$750 \kms).  For each of the radial velocities, we summed a spectral window (6 \AA\, wide) on each spaxel, centered on each line of the doublet.  Then we removed the underlying background emission in the window for each spaxel by subtracting the average of two 6 \AA\, windows on either side of the [S~II] doublet.  Finally we divided the 6716 \AA\/ by the 6731 \AA\/ image. This process was straightforward for the outermost (fastest ring), though it became more complicated for the low velocity ring ($\Delta$v = $-$750 \kms) due to blueshifting of the [S~II] $\lambda$6731 line from the low velocity ring to shorter wavelengths.  This resulted in blending of the 6731 \AA\, ejecta line with the background [S~II] $\lambda$6716 line.    The blending problem is illustrated in Figure~\ref{fig7}, where the [S~II] line profiles from one spaxel on the innermost ring are shown.  The background contribution had to be corrected before narrow band images of the inner ring could be extracted.  We performed the correction by first extracting a background spectrum from a 3$\times$3 spaxel box (3${\arcsec}\times$3${\arcsec}$) located just outside the inner ring.   We then subtracted that spectrum from each spaxel of the inner ring (in the lower panel of Figure~\ref{fig7} we show the result of subtracting this spectrum from the blended one).  We finally generated narrowband [S~II] images of the inner ring from the background subtracted data.   

The final [S~II] line ratio map was generated by adding the narrowband [S~II] $\lambda$6716 images of all three rings together, then doing the same for [S~II] $\lambda$6731, and finally dividing the summed images by one another.  The result is shown in Figure~\ref{fig8}.  Although the ratio image shows fine scale noisiness (especially in the faintest ring emission, where background subtraction adds additional noise), patterns can nevertheless be identified in the [S~II] line ratios.  The [S~II] emission from the outermost (N-rich) ring shows F($\lambda$6716)/F($\lambda$6731) $\sim$ 0.9-1.0, indicating $n_e\,\sim$700-1000 cm$^{-3}$ in the emission zone behind the shock.  The inner rings of the swirl show regions with F($\lambda$6716)/F($\lambda$6731) $\sim$ 1.3-1.4, indicating a lower density, $n_e\,\lesssim$70 cm$^{-3}$.  The reason for the higher average densities in what should be the outermost ejecta layers is unclear.  The higher densities may be due to the dynamical effects (such as Raleigh-Taylor instabilities) in the ejecta.  

\section{Kinematics of the Swirl}

The patterns in radial 
velocity and composition seen in the Swirl are broadly consistent with material from the hydrostatic layers of a massive star immediately before the supernova explosion (\cite{2003ApJ...591..288H}).  In particular, the N-rich outermost layer at $-$1350 km/s shows very little O or S emission, suggesting that it arises from the outermost CNO-processing layers of the core, possibly from a stripped progenitor.  The next layer around $-$1100 km/s shows strong O emission, possibly indicating this material represents the boundary between the CNO-processing layer and the hydrostatic burning layers of O, while the innermost layers around $-$750 km/s show strong S and Ar emission but weaker O emission, consistent with still deeper layers where O may have undergone explosive nucleosynthesis.  The presence of N emission at all velocities may indicate some moderate overturning and mixing during the explosion.

One thing clear from the kinematic maps of the Swirl is that it cannot be viewed as simply the result of limb brightening of expanding spherical shells, as in a normal supernova remnant.  Limb brightening occurs where the line of sight crosses a large path length through a narrow layer of emitting gas.  For a spherical shell, this occurs close to  tangency with the line of sight, implying that the gas motion at the edge of the shell should lie nearly in the plane of the sky.  The velocity component toward the observer should be low.  To the contrary, as we have shown, the rings exhibit a radial velocity of many hundreds of kilometers per second.  Although we have yet to measure  the proper motion of the Swirl to determine its velocity perpendicular to the line of sight, it is already evident that the Swirl traces a component of SN ejecta with a significant bulk motion toward us.   

The offset between ring centers can also explain the asymmetric morphology of the outermost (nitrogen-rich) ring.  As can be seen in the H$\alpha$+[N~II] image in Figure~\ref{fig:fig2}, the outermost ring deviates from a circular shape, and follows a roughly elliptical shape squashed in the E-W direction. This may indicate that it is not oriented completely in the plane of the sky.  It is clear that the eastern side of the ring has partially dissolved into finer knots, or what may even be termed a 'mist', while the western side looks more intact, though also clearly showing clumpy morphology.  These differences may be due to differences in where the reverse shock in Puppis A struck the rings.  If it struck the eastern side of the outermost ring first, then that side has had more time to become disrupted, while the other rings have been shocked more recently.  One possible picture of the rings is shown in the sketch in Figure~\ref{fig9}.  

The rings are not exactly circles, but we can obtain approximate sizes and positions within the remnant.  Assuming an age of 3700 years, line of sight velocities of -750, -1000 and -1350 km/s imply that the rings have moved 2.8, 3.7 and 5.0 pc toward the Earth from the original explosion site.  Assuming a distance of 2.2 kpc, the separation of about 8\farcm5\/ between the rings and the CCO indicates a relative velocity of about 1500 \kms.  The proper motion of the CCO is about 763$\pm$73 \kms\/ at a distance of 2.2 kpc \citep{mayer20} in a direction away from the rings, so  the centroids of the rings have been moving at about 750 \kms in the plane of the sky away from the explosion site.  The 1\farcm5\/ radius of the inner ring indicates that the diameter has been growing at 130 \kms, so the full angle of the conical expansion would be about 15$^\circ$.  The intermediate and outer ring diameters and positions are consistent with that opening angle.

\section{Swirl Composition}

We can obtain a crude mass estimate for the rings by first utilizing radiative shock models to match emission line fluxes for different element species.  We can then use the predicted abundances to calculate the total mass of the Swirl subject to assumptions on the geometry and physical sizes.

The shock speed, preshock density and composition of the rings can in principle be modeled by comparing line strengths in the WiFeS spectra with predictions from radiative shock models (\citealt{raymond79}).  There are complicating factors, however.   A planar, steady flow shock model is a very crude approximation to what is likely a much more complex flow.   For example, the fragmentation of the rings into smaller clumps can clearly be seen in the archival\footnote{This research uses services or data provided by the Astro Data Archive at NSF's NOIRLab. NOIRLab is operated by the Association of Universities for Research in Astronomy (AURA), Inc. under a cooperative agreement with the National Science Foundation.}  H$\alpha$ image of the Swirl (Figure \ref{fig:fig2}), as well as the H$\alpha$ and [N~II] images of the Swirl from \wifes (Figure \ref{fig3}).  In particular, the observed line widths are mostly broader than the instrumental resolution and are of order 150 \kms, indicating strong turbulence.  It is likely that when the dense ejecta encounter the reverse shock, Kelvin-Helmholtz instabilities in the shear layer between the high and low density gas will create turbulence and hence a range of shock speeds.  Moreover, the expansion time of the shocked gas is roughly equal to the SNR age, which in turn is comparable to the recombination time scale, so that an expansion cooling term might be significant though it is not included.

Nevertheless, we will compare 1D steady flow shock models with the observed spectra under the assumption that if a model matches the density (as determined from the [S II] ratio), relative ionization fractions of N and O, and temperature (as determined from [O III] I($\lambda$4363)/I($\lambda$5007) or [N II] I($\lambda$5755)/I($\lambda$6548), then it matches the average conditions in the emitting region.  Therefore, the predicted relative intensities of lines of different elements can be used to estimate the abundance ratios.  Shocks in highly enriched SN ejecta can behave much differently that shocks in hydrogen-dominated plasmas \citep{raymond18} because of the enormous radiative cooling rates.  We use the shock model of \citep{raymond79} as modified for shocks in highly enriched ejecta as employed by \citep{blair00}.  It allows for unequal electron and ion temperatures and essentially pure heavy elements, but it does not include thermal conduction \citep{borkowski90}, which might or might not be suppressed by magnetic fields.

\section{Shock Models of the Swirl }

Although the full list of spectral lines observed in the Swirl is not exceptionally long, there is substantial spatial variation in line strengths of emission lines.  Ideally the most robust constraints on shock parameters are obtained by modeling spectra exhibiting a wide a range of ionization stages from a wide range of elemental species.  Unfortunately this is difficult for the outermost (fastest) ring material, which is dominated by nitrogen line emission and shows weak sulfur emission and little to no emission from hydrogen, oxygen or other species.  Slower material in the inner rings is similarly problematic, with localized regions showing only a range of ionization stages from a limited number of elemental species ([O~I] - [O~III]).  The intermediate velocity ring ($\sim$-1000 kms/) is one of the brightest segments, and shows a wider range of emission lines than most other parts of the ring, making it most amenable to modeling.  This segment, which we labeled P1, is marked by the arrows in Figure \ref{fig3}). 

Before modeling the P1 spectrum, we corrected the data for interstellar reddening.  First we gauged the extinction parameter E(B\,-\,V) by using the relationship between optical extinction ($A_V$) and hydrogen column ($N_H$) derived by \citep{2009MNRAS.400.2050G}.  Taking $A_V\,\approx$1.58 \citep{2011ApJ...737..103S} and $N_H$ = 3$\times$10$^{21}$ cm$^{-2}$ from the X-ray analysis of Puppis A \citep{2010ApJ...714.1725K} we found  E(B\,-\,V) = 0.4.  While E(B\,-\,V) values ranging from 0.2 \citep{1995ApJ...454L..35B} up to as high as 0.7 \citep{2017MNRAS.464.3029R} have been estimated in prior studies of Puppis A, the ratio most likely exhibits some variation across the face of Puppis A, and our value is a plausible one for the central region.  Combining this E(B\,-\,V) with the  prescription of \cite{1999PASP..111...63F}, we corrected both the red and blue channel spectra of P1 for interstellar reddening.   

Upon inspecting the corrected line fluxes, we found a Balmer decrement F(H$\alpha$)/F(H$\beta$) close to 1.5 for the ejecta, which is about half the value expected for hydrogen recombination in a 10,000-20,000~K gas.  We found a similar problematic ratio for the diffuse background emission just outside the Swirl, where narrow Balmer line emission (H$\gamma$, H$\delta$ and H$\epsilon$) is detected. The ratios of these higher Balmer lines relative to H$\beta$ in the background spectrum disagree even more strongly with theoretical predictions.  This is mainly due to inaccuracy in the blue channel flux calibration mentioned earlier, which also resulted in flux miscalibration between the red and blue channels.  We used the background hydrogen lines to calculate a wavelength dependent correction factor for emission lines shortward of 4000 \AA.  We assumed that this background, seen toward the projected interior of Puppis A, lies at nearly the same distance as the SNR and originates in circumstellar gas that was photoionized by either the massive progenitor (possibly a WN-type progenitor; \citealt{1983prhe.work..249D}) or by the UV flash during shock breakout. As such, this gas is at nearly the same distance as Puppis A and so the extinction correction for Puppis A should also be applicable to these lines.    

To correct the blue channel emission line fluxes, we calculated the wavelength-dependent scaling factors needed to correct each of the dereddened Balmer lines from H$\beta$ down to H$\epsilon$ to their theoretically predicted ratios relative to H$\alpha$.  First we verified that the red channel was properly flux calibrated and dereddened.  Then we computed the multiplicative factor needed to  correct the H$\beta$ flux and match the theoretically predicted Balmer decrement (2.86 for a 10,000~K recombining gas).   Next, we calculated the correction factors needed to bring the ratios of the higher Balmer lines H$\gamma$, H$\delta$ and H$\epsilon$ into agreement with their theoretically predicted ratios relative to H$\beta$ \citep{2011piim.book.....D}.   These correction factors ranged from around 10 at 3970 \AA\, to around 2.7 at H$\gamma$.  Finally we interpolated from these correction factors to adjust the intensities of all lines in the blue band.  
The final dereddened surface brightnesses are shown in Table \ref{tab1}, along with results from a single radiative shock model matching the observations.  The preshock ionization state is unclear, due to our uncertainty in the ionizing radiation field produced by each shock.  As described above, the completeness of the recombination zone is uncertain, so equilibrium pre-ionization cannot be assumed.  

In modeling Position P1 in Table 1, we use the shock model code of \citet{raymond79} with updated atomic rates from \citet{1987ApJ...316..323H}. The line centroids are all near -1000 \kms , so the emission originates from the intermediate ring.  We present one model for comparison, with the caveat that the parameters are not unique.  Since emission lines from only a modest number of elements are detected over a narrow range in ionization state, the modeling involves a trade off in model parameters (shock speed against oxygen abundance, for instance).  The very strong nitrogen emission and weak oxygen lines suggest that the gas has undergone CNO processing, in which H is converted to He, and C and O are converted to N \citep{1995ApJS..101..181W}.  We chose an abundance set in which most of the H has been converted to He, which increases the abundances of all the elements relative to H, including those that do not take part in the CNO process.  The comparison model uses a shock speed of 80 \kms, preshock density of 8 $\rm cm^{-3}$, 90\% preionization of H and He and a magnetic field of 0.1 $\mu$G (a weak magnetic field consistent with the supernova ejecta origin for the material).  The logarithmic abundances are  (H\,:\,He\,:\,C\,:\,N\,:\,O\,:\,Ne\,:\,Mg\,:\,Si\,:\,S\,:\,Ar\,:\,Ca\,:\,Fe) =  (12.0\,:\,11.4\,:\,8.2\,:\,9.6\,:\,8.0\,:\,8.3\,:\,8.0\,:\,7.9\,:\,8.0\,:\,7.7\,:\,6.9\,:\,7.0\,:\,7.7).  We note that the high brightness of the observed [Ca II] $\lambda$7291 line compared to the model predictions is almost certainly due to the line being pumped by the Ca H and K lines, a process that is not included in the model.  We also note that we have no real constraints on the C, Mg or Si abundances.

Although the models match the observations rather well, it is important to keep in mind that our models are steady-flow, 1D models, and the actual shocks are likely to violate these conditions to varying degrees.   The [S II] ratios shown in Figure~\ref{fig8} indicate that most of the outermost ring and substantial parts of the other two rings are in the low density limit.  The recombination time of hydrogen is about 80,000/n$_e$ years, so in those positions the recombination time is 1000 years or more.  The rings  are a small fraction of the current radius of the remnant, so they would have encountered the reverse shock at some fraction of the 3700 year age of the remnant, perhaps 1000 to 2000 years ago.  A second potential problem is that for very low densities, the cooling time may be comparable to the expansion time of the SNR, and the models do not include an expansion cooling term.  The densities in the outer ring are low enough that we do not attempt to match the fluxes to a shock model.  The [S II] ratio at position P1 indicates a density of 150-200 $\rm cm^{-3}$ according to CHIANTI \citep{dere97, delzanna15}.  Although we use the 1D shock model, it is possible that the shock is 'incomplete' in the sense that the gas has not had time to fully cool and recombine \citep{raymond88}.  Overall, while the model parameters are not unique, the conclusion that He and N are very overabundant compared to H ($\sim$2.6 times and 35 times solar), and that O is underabundant ($\sim$ 1/10th solar) is secure.  

In some locations in the Swirl a spatially layered pattern of emission can be seen from multiple ionization stages of the same element.  For the slow ring ($-$750 km/s) this is seen in the emission lines of [O~I], [O~II] and [O~III] (Figure \ref{fig10}).  The eastern side of the ring shows superimposed emission from all three species, while the western side shows [O~I] and [O~II] in a thin layer on the inner edge, with considerably more extended [O~III] emission outside of that.  This raises the possibility that all these layers may be consistent with a single spatially resolved shock.  In Table~2 we present line intensities for a spectrum (shown in Figure \ref{fig6}) taken from a cross cut through the intermediate velocity ring (marked by the white bar in Figure \ref{fig10}).  

Due to the apparently extreme overabundance of oxygen relative to hydrogen required to match the surface brightnesses in Table~2, we modeled the intensities for these shocks using a code designed for shocks in SN ejecta.  The model we used assumes a shock speed of 80 \kms, a preshock density of 3 cm$^{-3}$ and a magnetic field of 0.1 $\mu$G, and it matches the [S II] density-sensitive ratio.  The measured value of the [O II] density-sensitive ratio lies above the low density limit, so we rely on [S II] for the density.  The gas was assumed to be 50\% ionized with relative abundances (H:He:N:O:Ne:S) = (12:12.9:11.1:11.3:10.9:10.1).  The electron and ion temperatures were taken to be nearly equal at the shock based on observations of nearly full electron-ion equilibration in relatively slow shocks in the Cygnus Loop \citep{ghavamian01, salvesen09, medina14}.  However, the electrons cool very rapidly by collisionally exciting ions, and their temperature quickly falls to about 1/4 the ion temperature.  Only after the electrons cool to about 50,000 K can Coulomb collisions bring the electrons and ions toward equilibrium.

Table~2 shows that most of the predicted line ratios match the observations, with the major exception of [O I] $\lambda$6300.  This is a longstanding problem with the interpretation of shocks in SN ejecta such as the Fast Moving Knots in Cas A or the O-rich regions of N132D and 1E102.2$-$7219 \citep{itoh88, 2000ApJ...537..667B, raymond18}.  The difficulty is that the gas that is ionized to produce the [O III] emission cools so fast that it cannot recombine to form [O I].  In fact, the model in Table 2 was finely tuned to produce [O II] and [O III] emission in the observed proportions, because even slightly higher shock speeds produce very little [O II].  The usual explanation is that a range of shock speeds is present in even a small region of space due to either preexisting small scale density structure or to shear instabilities that shred the ejecta knots \citep{klein94}.  

Based on the above arguments, a 1D model is an extremely crude approximation to the actual conditions.  However, the overall ionization fractions of the singly- and doubly-ionized states should be approximately correct, so that we can infer the Ne/O abundance ratio from [Ne III]/[O III], and the N/O and S/O ratios can be estimated from [N II]/[O II] and [S II]/[O II].  The He/H ratio would in principle be easy to infer from the ratio of the 5876 \AA\/ line to H$\alpha$, but that line ratio is also sensitive to the shock speed and the preshock ionization fraction.  Moreover, the He~I $\lambda$5876\AA\/ line centroid at -910 \kms is significantly offset from the -700 to -800 \kms centroids of most of the other lines, so it may originate from a different place along the line of sight.  The O/H abundance ratio can be inferred from the ratios of the [O II] lines to the Balmer lines, because the Balmer lines are produced by recombination of H II, and the H II ionization fraction is tightly tied to that of O II by charge transfer.  The [O II] emissivity is somewhat sensitive to temperature, but the agreement between observed and predicted ratios of the 7320\AA\/ multiplet to the 3727\AA\/ doublet indicates that the temperature in the [O II] emission zone is approximately correct.

With the above caveats, we take the abundance ratios of the model to be correct to within a factor of 2, with the exception of the uncertainty in the He abundance, which is larger $--$ perhaps a factor of 4. It is notable that there is a substantial amount of hydrogen, but that burning of H to He can account for some of the high ratios of the other elements to H.  Although the strongest emission lines from the rings  follow a pattern suggestive of progressively deeper layers of the star (as  seen in Figure~\ref{fig5}), the specific abundance ratios do not seem to match any layer in an exploding star.  Therefore, it is possible that different layers of the star were mixed on small scales during the explosion, or even that material from the exploding star mixed with material from a binary companion.  In either case, the mixing may have occurred through the process that formed the ring structure.  By comparison, the Fast Moving Knots in Cas A are devoid of hydrogen and nitrogen, and they tend to be strongly dominated by O or by S, but not a mixture (see \citealt{2006ApJ...636..859F} and references therein).

We can estimate the mass of the rings using their dereddened surface brightness, their measured angular sizes and the shock model results presented above.  The surface brightness $F$ (ergs cm$^{-2}$ s$^{-1}$ sr$^{-1}$) of a region of thickness $d$ is
\begin{equation}
    F\,=\,\frac{h\nu}{4\pi} n_e\, n_i\, q(T)\, d
\end{equation}
where $q(T)$ is the excitation coefficient of the emission line in question at temperature T, $\nu$ is the frequency of the emission line, and $n_{e,i}$ are the densities of the electrons and ions, respectively.  We use an average electron density $n_e\,\approx\,$100 cm$^{-3}$ as found from the [S~II] line ratio of region P1.  Most of the mass of the emitting rings is contained in ions, with $n_i\,\approx\,M / \mu m_H V$, where $\mu$ is the mean mass per particle and $V$ the emission volume.  Substituting in the relation above and solving for $M$:

\begin{equation}
    M\,=\,\frac{4\pi}{h\nu}\left(\frac{\mu\, m_H\, V}{q(T)\,d}\right)\,\left(\frac{F}{n_e}\right)
\end{equation}

where $V$ is the volume of the emitting region (= $A\,L$, where $A$ is the ring area on the sky and $L$ is the line of sight ring depth).  $L$ is uncertain, but if we assume it equals the postshock thickness $d$ of an 80 km/s shock in medium of density 8 cm$^{-3}$ (as estimated from our model for P1), then the factors of $d$ cancel in the top and bottom of the equation above.  We can take the area of the emitting rings as $A\,=\,2\pi \theta \Delta \theta\,D^2$, where $\theta$ and $\Delta \theta$ are the ring angular radius and width, and $D$ is the distance to Puppis A.  Then the above relation becomes

\begin{equation}
    M\,=\,\frac{8\pi^2 D^2}{h\nu}\left(\frac{\mu\, m_H  }{q(T)\,}\right)\,\left(\frac{\, \theta \Delta \theta}{n_e}\right)\,F
\end{equation}

Based on our shock models, $\mu\,\approx$\,2.08 for the derived abundances.  The outermost ring ($-$1350 km/s) can be approximated as a half annulus of width 20$^{\arcsec}$ and outer semiminor and semimajor axes of 80$^{\arcsec}$ and of 105$^{\arcsec}$.  Using the dereddened H$\alpha$ surface brightness from P1 in Table~1 as a typical value, assuming D = 2.2 kpc and taking $q_{H\alpha}(10,000~K)\,\approx\,$10$^{-13}$ cm$^{3}$ s$^{-1}$ (\citealt{2011piim.book.....D}) yields a mass of approximately 0.16 $M_{\odot}$ for the outermost ring.  The  inner two rings likely have masses comparable to these values.  The distance to Puppis A is the largest source of uncertainty in this estimate.  If we include both this source, as well as a 20\% uncertainty due to variation in composition between rings, we crudely estimate 0.05 $\lesssim$ M $\lesssim$0.5 $M_{\odot}$ for the entire Swirl.  

\section{Discussion}

\begin{figure}[t] 
 \centering
\includegraphics[width=7in]{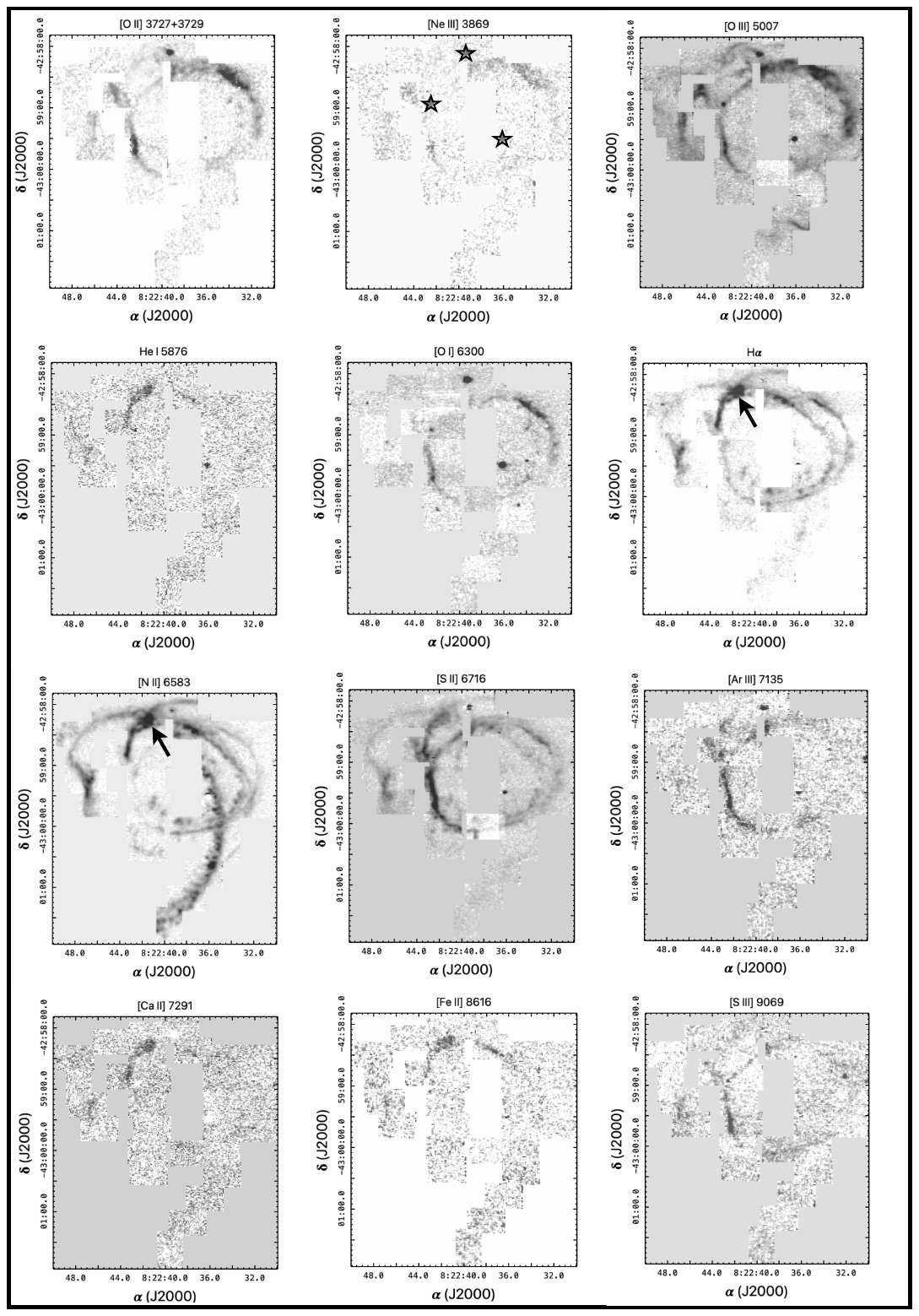}  
   \caption{Narrow band images of the Swirl extracted from most detectable emission lines in the WiFeS datacube.  Each image has been continuum subtracted, though leaving stellar residuals from three stars (marked by the star markers in the [Ne~III] image).  The arrows point to one of the spectral extraction locations (P1) used in shock modeling (fluxes shown in Table~1).   }
   \label{fig3}
\end{figure}

\begin{figure}[ht] 
 \centering
\includegraphics[width=6in]{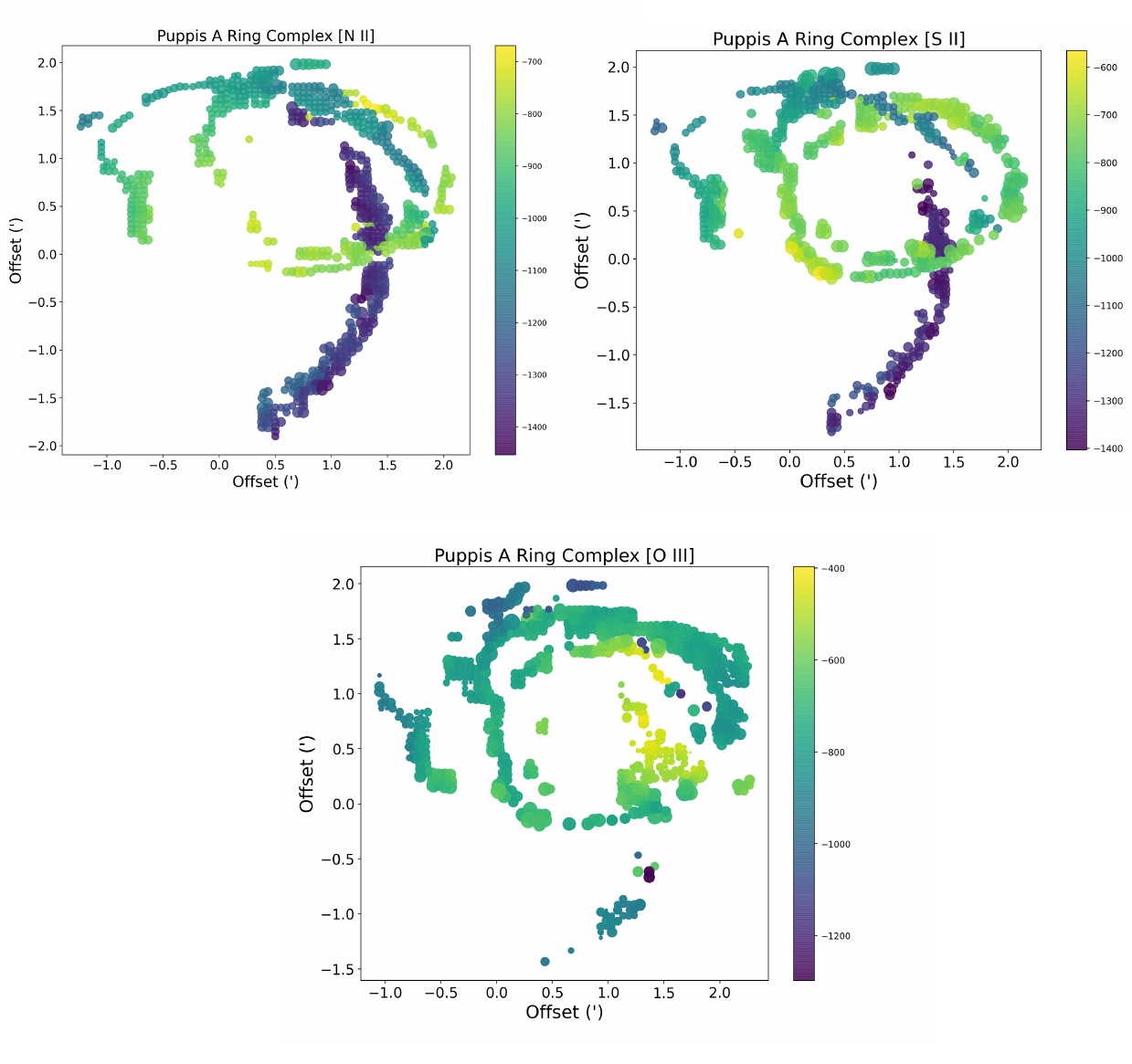}  
   \caption{ Radial velocity maps (in km/s) of the Swirl in Puppis A.  The width of each data point reflects the Doppler width of each fitted emission line, with the smallest dots equal to the unresolved instrumental width (110 \kms).  North is up and East is toward the left (as in the right panel of Figure 1).  The coordinates in the image are defined as offsets from the approximate center of the outermost ring (located at ($\alpha$, $\delta$) = (08:22:42.67, $-$42:59:55.1)) }
   \label{fig4}
\end{figure}

\begin{figure}[ht] 
 \centering
\includegraphics[width=5in]{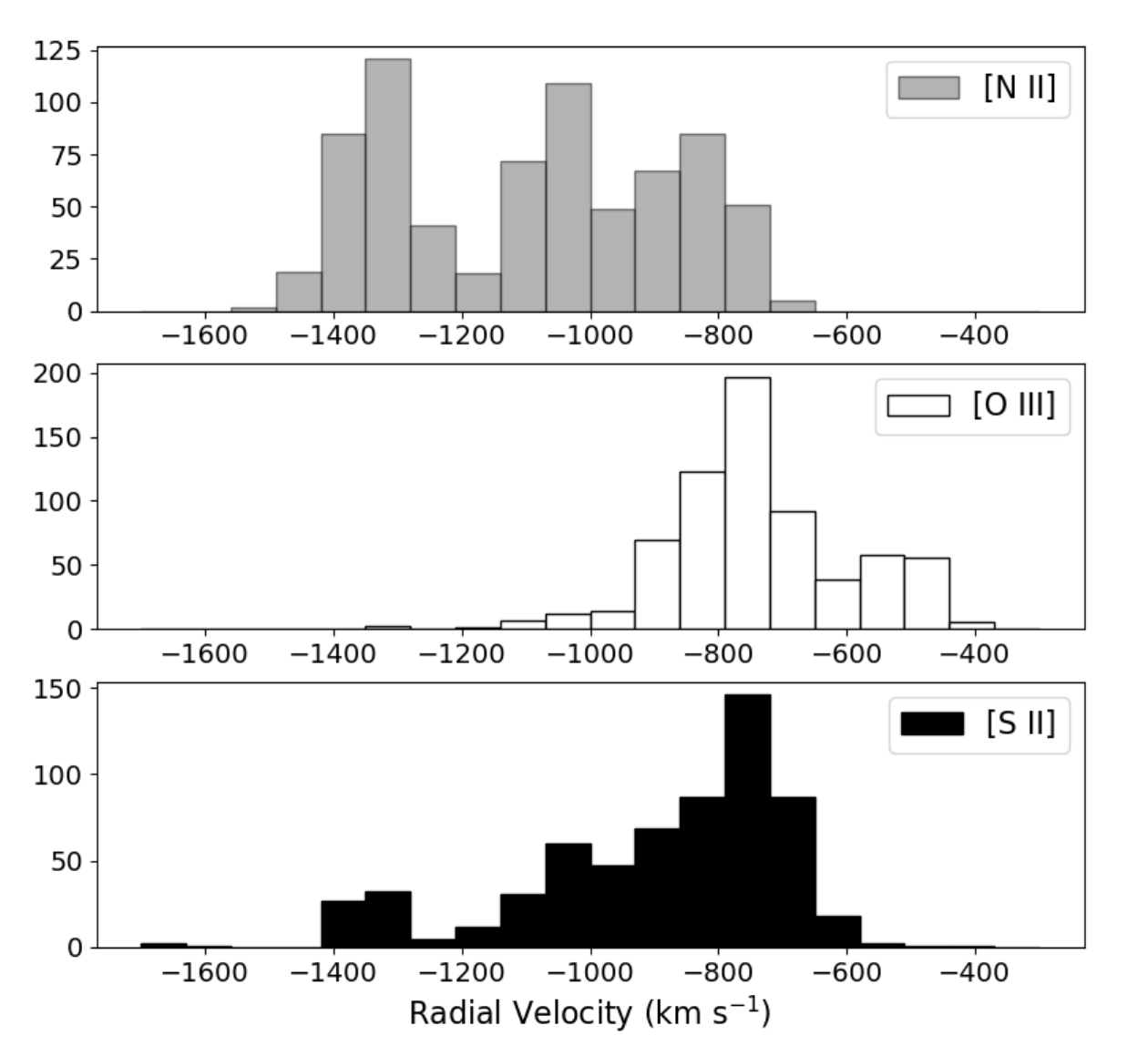}  
   \caption{Histogram of the radial velocities obtained by fitting spectral line profiles in [N~II] $\lambda$ 6548, [O~III] $\lambda$5007 and [S~II] $\lambda$6716 from the Swirl.  Three peaks can be identified in the [N~II] velocity distribution around $|\Delta v|$ = 1325 km/s, 1025 km/s and 750 km/s, while the [O~III] and [S~II] emission is concentrated deeper within the slower moving ejecta, at $|\Delta v|$ = 750 km/s.    }
   \label{fig5}
\end{figure}

\begin{figure}[ht] 
 \centering
\includegraphics[width=7in]{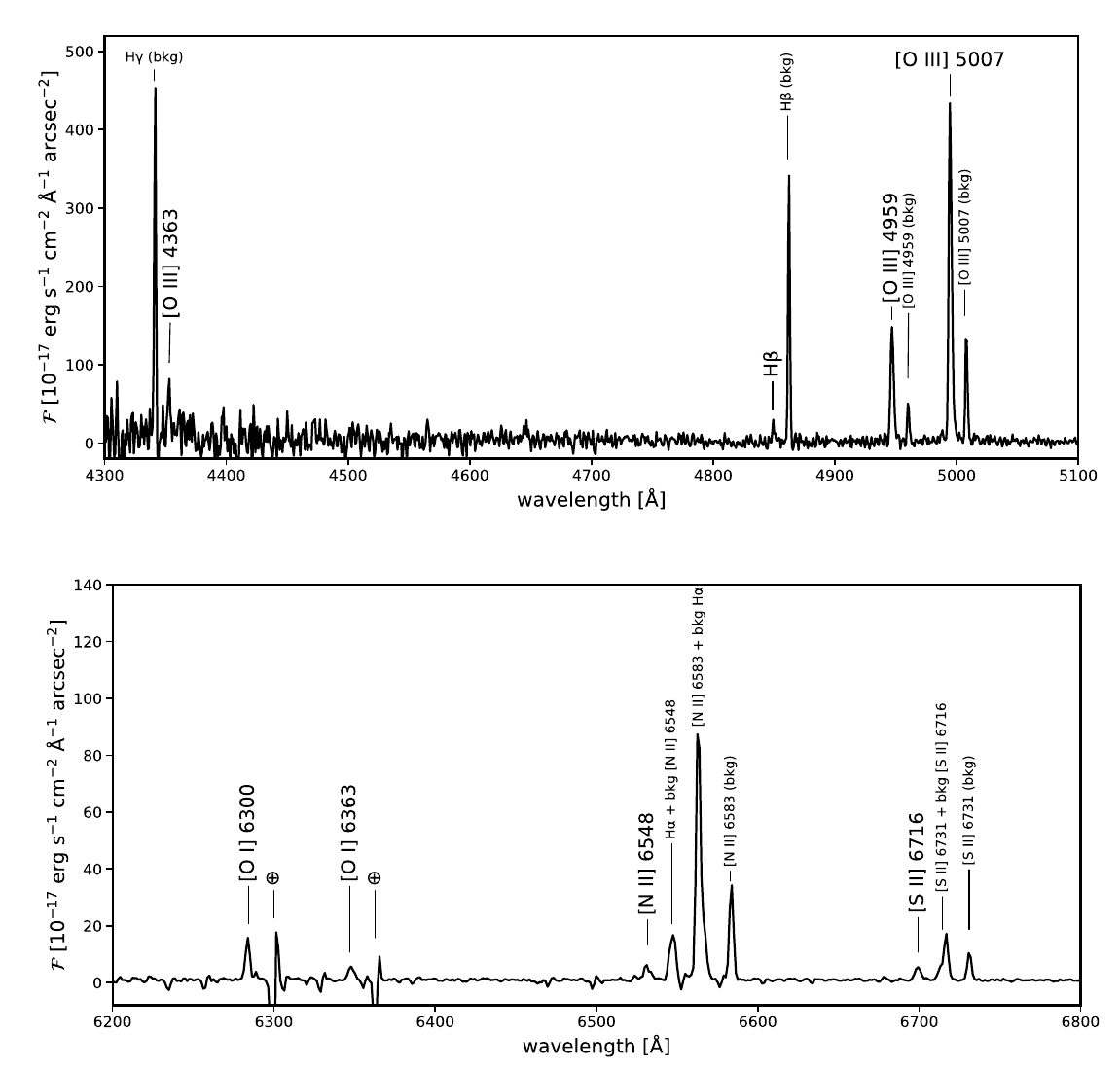}  
   \caption{The dereddened specific flux per square arcsecond from the crosscut through the Puppis A rings marked in Figure~10.  Prominent emission lines from the ejecta are marked.  Those from background nebular emission (bkg) are also labeled.
    }
   \label{fig6}
\end{figure}

\begin{figure}[ht] 
 \centering
\includegraphics[width=5in]{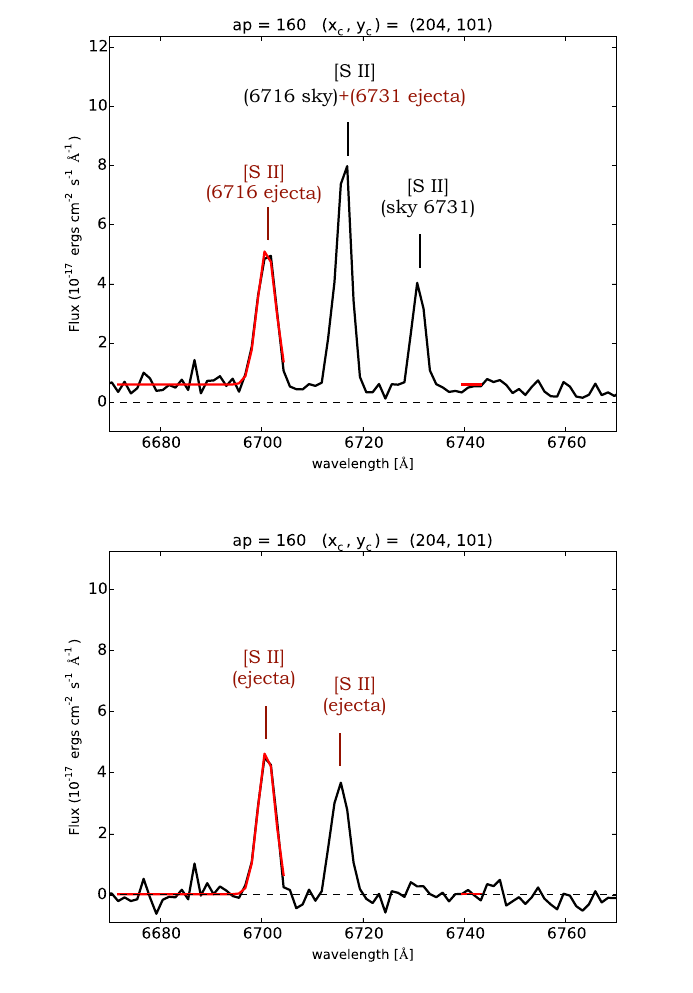}  
   \caption{ Top: Example of a [S~II] line profile where ejecta [S~II]$\lambda$6731 emission is Doppler shifted to blend with background [S~II]$\lambda$6716 emission.   The red line indicates the best profile fit for the 6716 \AA\, line (used in measuring radial velocities).   Bottom: the same spectrum but with a 1-dimensional sky spectrum (including nebular emission) subtracted, leaving only the ejecta [S~II] lines.    }
   \label{fig7}
\end{figure}

\begin{figure}[ht] 
 \centering
\includegraphics[width=6in]{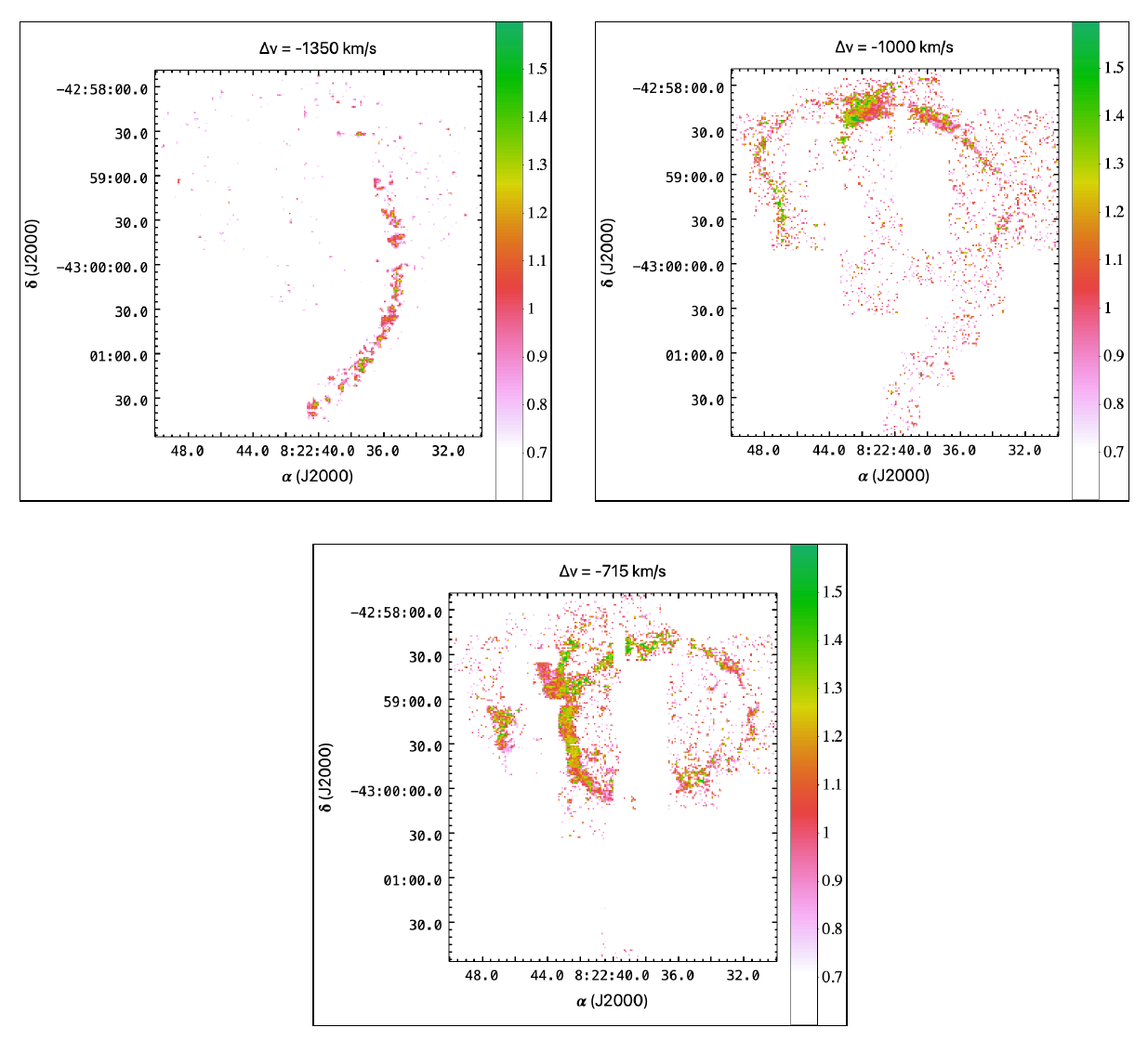}  
   \caption{ Ratio map of the ejecta rings in Puppis A for [S II] $\lambda$6716/6731.  Much of the ring material shows [S~II] ratios near the low density limit (corresponding with fainter emission), while some regions having enhanced densities (F($\lambda$6716)/F($\lambda$6731)$\lesssim$\,1.0) correspond with brighter emission.      }
   \label{fig8}
\end{figure}

\begin{figure}[ht] 
 \centering
\includegraphics[width=6in]{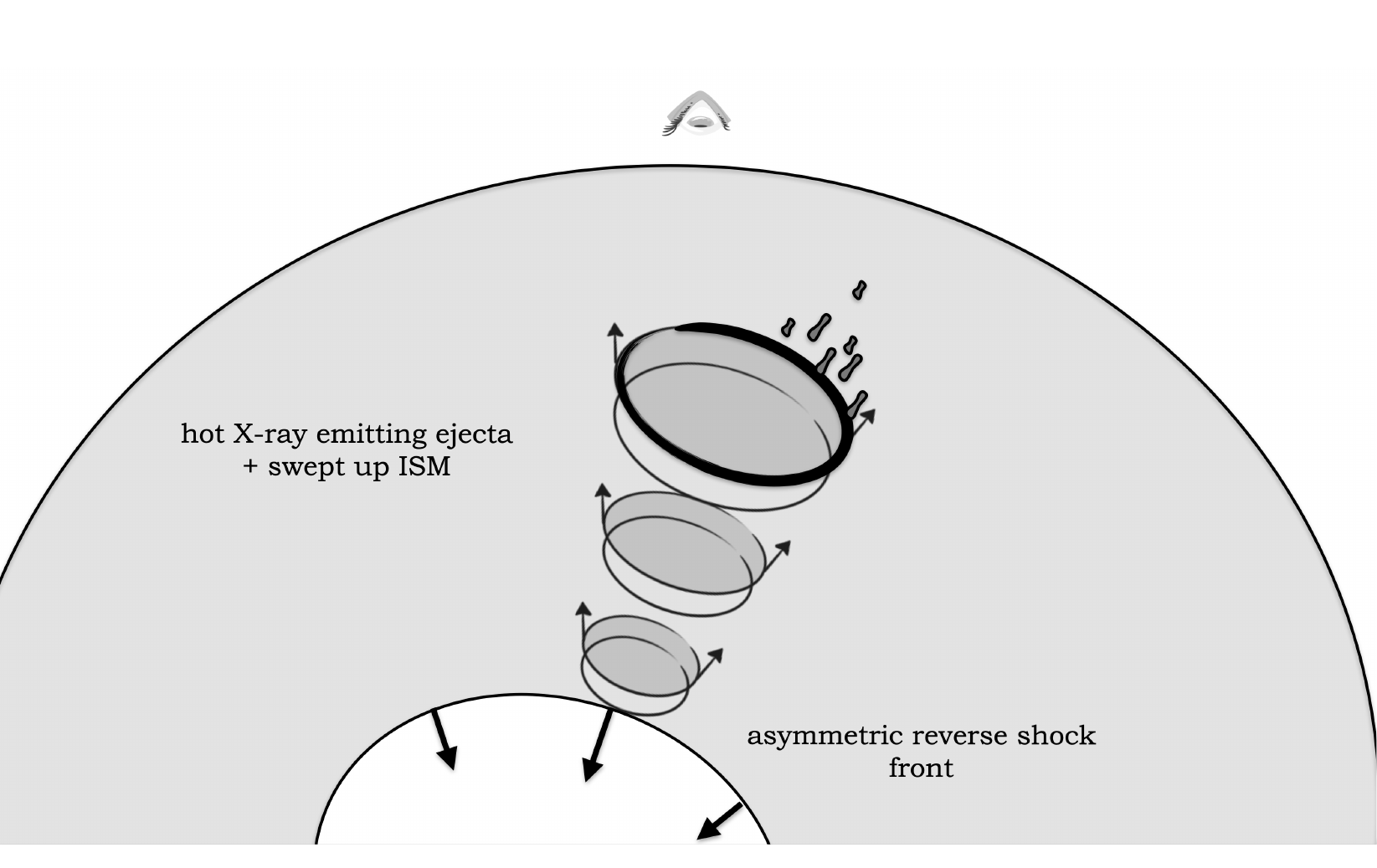}  
   \caption{ Sketch of the ring structure and its location relative to the reverse shock and swept up ISM gas.  The viewing location is marked at the top.  The largest ring (corresponding to the outermost layers of the exploded star), having encountered the reverse shock in the most distant past, has fragmented the most as it is disrupted by internal shocks.    }
   \label{fig9}
\end{figure}

\begin{figure}[ht] 
 \centering
\includegraphics[width=5in]{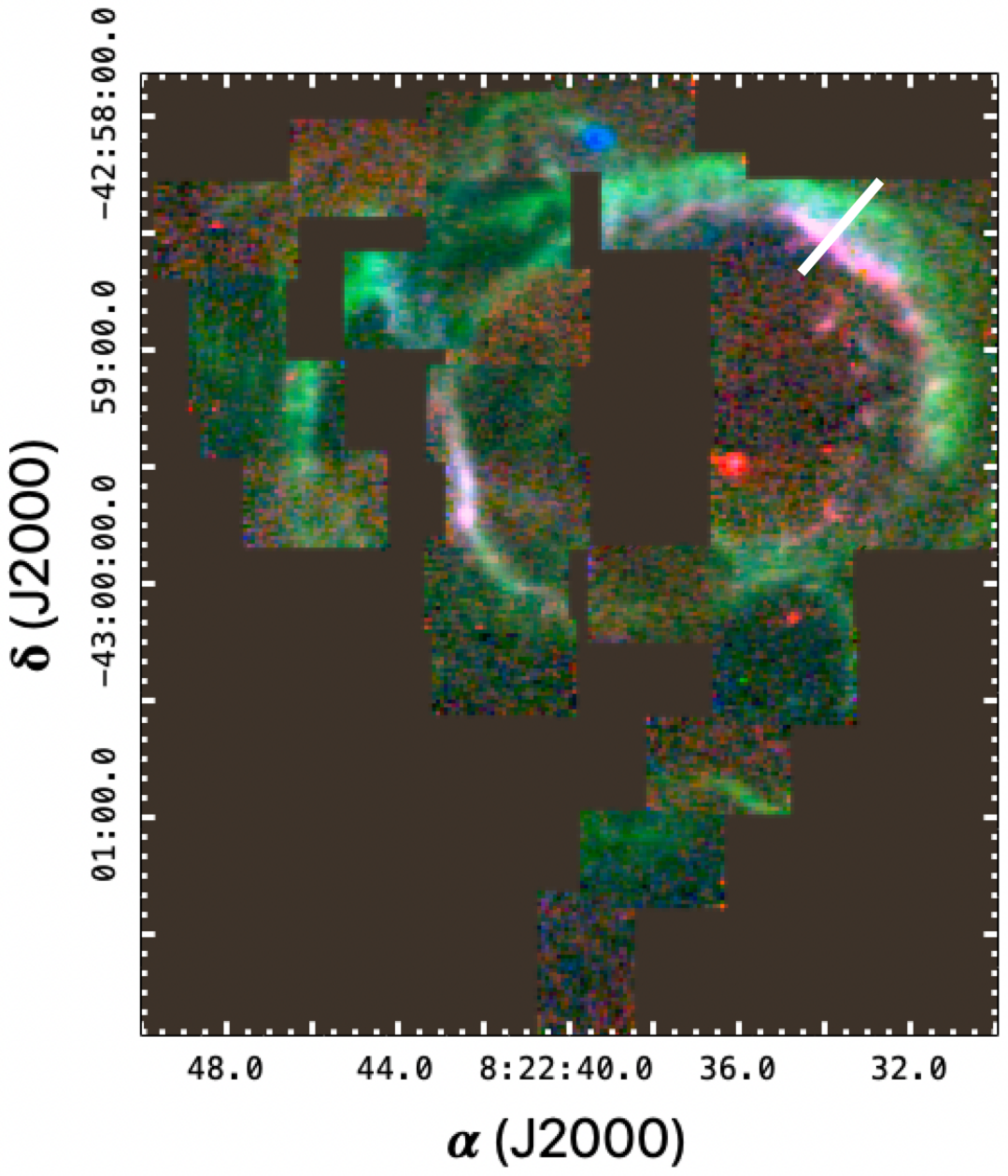}  
   \caption{ Color image of the Swirl at a radial velocity of $-$750 km/s, showing emission from [O~I] (red), [O~III] (green) and [O~II] (blue).  The spectral extraction window of the inner ring used for the shock modeling in Table~2 is marked by the white rectangle.   }
   \label{fig10}
\end{figure}

The first question to consider is whether the Swirl is even related to the X-ray remnant Puppis A.  The similarity of the Swirl's radial velocities and composition to those of other O-rich features known in Puppis A (such as the Omega filament and other O-rich ejecta; \citealt{1989Natur.337...48W}, \citealt{1995ApJ...439..381S}), as well as its location near the center of Puppis A make it very likely that the Swirl and the X-ray remnant are related.   Also relevant are that earlier spectroscopic observations by \cite{1977ApJS...33..437D} and \cite{1995ApJ...439..365S}, who found elevated nitrogen abundances in slow-moving ($\sim$140 km/s) clumpy circumstellar material on the western side of Puppis A.  
\vskip 0.2in
The next question is whether the Swirl is circumstellar wind material from the progenitor, or supernova ejecta. Properties of the Swirl which favor a supernova origin are:

\begin{enumerate}

\item The radial velocities of the Swirl complex  measured from the \wifes data greatly exceed those expected for clumpy stellar wind material, as seen for example in SN 1987A ($\sim$ 20 km/s).   

\item Although the Swirl has an ordered motion reminiscent of a wide jet, its kinematics appear inconsistent with a jet as seen in e.g., Cassiopeia A, \cite{2001AJ....122.2644F}.  Unlike Cas A, the Swirl has no discernible redshifted counterpart, and furthermore its maximum velocity does not exceed 1350 km/s.  This is around the same speed as other O-rich ejecta knots seen in Puppis A, and is in fact even slower than the values measured in the $\Omega$ filament (1500 \kms; \citealt{1985ApJ...299..981W}; \citealt{1988srim.conf...65W}).  

\item The Swirl rings show a progression of heavy element composition consistent with layers from hydrostatic burning on a massive star.  The largest ring is also the fastest ($-$1350 \kms) and is dominated by [N II] emission, with weak lines from other species.  This likely originates in the outer layers of a red supergiant, where H has been lost to either a wind or gravitational pull of a binary companion. The inner rings also show strong emission from heavier species such as O and S, but are also accompanied by strong emission from N.  These are consistent with progressively deeper, slower moving layers of supernova ejecta.  The presence of N in the slower moving ejecta may be evidence of mixing between the inner and outer layers of the star during the explosion.  Note that one compositional peculiarity of the rings is the enhanced emission in [Fe~II] $\lambda$8616 and [Ca~II] $\lambda$7291 seen in the intermediate velocity ring (Figure~\ref{fig3}).  Although not shown in that figure, [Ni~II] $\lambda$7378 is spatially co-located with the Ca and Fe emission.  Naively we would expect heavy species such as Ca, Fe and Ni to be found in the innermost, slowest ejecta ring ($\Delta$v = -750 \kms).  Their presence in the faster material may be evidence of outward mixing of this material from explosive nucleosynthesis. 
 
\item The average radii of the rings (0\farcm94, 1\farcm22 and 1\farcm4) correspond to radial velocities of ($-$750 km/s, $-$1100 km/s, $-$1350 km/s) $-$ a nearly linear relationship between radius and radial velocity.  This is consistent with a funnel-like expansion from a localized point.        

\item The Swirl kinematics are are not easily explained with a limb brightened spherical shell model.  The edges of such shells should have velocity vectors lying nearly in the plane of the sky, with a radial velocity much lower than  observed.   
\end{enumerate}

Based on these properties, we suggest that the kinematics, composition and morphology of the Swirl in Puppis A are most consistent with a 'funnel' (or shadow) carved out of heavy element ejecta expelled by the core collapse SN that produced Puppis A (most likely Type IIb/L,   \citealt{2005ApJ...619..839C}).   This shadow may be formed by an obstacle blocking the spherically expanding ejecta very close to its source.  Such an obstacle would be readily available in a stellar companion to the exploding star.  \cite{2020MNRAS.499.1154H} proposed just such a scenario in their models of lonely stripped-envelope supernovae.  In their models, the two would both be red supergiants when the slightly more massive one exploded.  Furthermore, they would be far enough apart to have evolved as essentially single stars, but close enough ($\sim$ 1000-4000 R$_{\odot}$) for a significant part of the secondary's enveloped to have been stripped off during the explosion of the primary.  They estimated that on the order of 0.3\%-1\% of core collapse SNe would be of this type, making a scenario such as that of Puppis A exceedingly rare.


Although it is appealing to invoke binarity of the progenitor system in explaining the Puppis A rings, this raises the obvious question: is there a detectable surviving companion?  If the scenario of \cite{2020MNRAS.499.1154H} applies, the outer envelope of the secondary could have been largely stripped off, leaving the surviving star in an altered state from an RSG.  There have in fact been at least two recent efforts to search for a binary companion in Puppis A using the newly available data from the Gaia EDR3.  \cite{2021AN....342..553L} and \cite{2021MNRAS.507.5832K} extracted proper motions for stars near the center of Puppis A and found no evidence of an unbound binary companion.  Indeed, using statistical searches for surviving binaries in additional supernova remnants, \cite{2021MNRAS.507.5832K} estimated that nearly 70\% of core collapse supernovae are unlikely to be binary systems at the time of explosion, with the remaining fraction divided nearly evenly between bound and unbound systems.  The CCO in Puppis A has no known optical counterpart, so a surviving neutron star-nondegenerate binary can also be ruled out.  By estimates of \cite{2021MNRAS.507.5832K} this would leave approximately a 13\% chance of an unbound binary outcome.  However, this is still not zero.  Furthermore, earlier studies  \citep{2021AN....342..553L, 2021MNRAS.507.5832K} utilized a search area centered on the geometric center of Puppis A, well separated from the location of the Swirl.  A larger search area may be necessary.

 
 Based on the results described above, searches for surviving binary companions have been performed in recent years for  Cas A \citep{2018MNRAS.473.1633K, 2019A&A...623A..34K}.  These searches have all failed to detect a surviving stellar companion, which poses a problem for the binary progenitor scenario.  However, in the lonely stripped envelope scenario of \citep{2020MNRAS.499.1154H} for Cas A they account for the lack of surviving companion by positing a binary system where one star was slightly more massive ($\lesssim$10\%) than the other.  They suggested that when the two stars became RSGs, the more massive one transferred mass to the secondary star and then exploded as a supernova around 10$^6$ years ago, stripping the outer envelope of the secondary and ejecting the newly formed neutron star.  After 10$^5$ or 10$^6$ years elapsed, the second star then went supernova, leaving the CCO seen today at the center of Cas A.  Observational evidence in nominal support of this scenario has been found in X-ray observations of the stellar wind material of Cas A \citep{2022A&A...666A...2O}.  
 
 A lonely stripped-envelope scenario sounds appealing for Puppis A, especially since it includes stripping of material from the secondary and formation of an ejecta cone similar to what is seen in the Swirl.  However, an ejecta cone should be a short-lived structure expected to dissipate on a time scale of a few thousand years as it is overrun by the reverse shock.  In the context of the \cite{2020MNRAS.499.1154H} model, the presence of recently shocked supernova remnant ejecta in the Swirl means that unlike Cas A, we may be seeing the remnant of the first SN rather than the second.    
 
 In their hydrodynamic models of lonely stripped-envelope supernovae, \citep{2020MNRAS.499.1154H} estimated that the secondary is shorn of nearly half its mass and its envelope is heated by the impact of the supernova shock to high temperatures, becoming blue in color for a few thousand years after the explosion.  The mass of the Puppis A Swirl is significant ($\lesssim$0.5 M$_{\odot}$), though we do not know how much of the Swirl's mass has been dissipated so far by the action of the reverse shock. Further studies are needed to determine how much can be attributed to the companion star and how much to the supernova itself.   
 

\begin{acknowledgments}

We thank Ashley Ruiter and Frank Winkler for their contributions to the observing proposals and discussions. The work of P. G. was supported by grants HST-GO-12545.08 and HST-GO-14359.011.  P. G. would also like to thank the University of New South Wales Canberra for the Rector's funded visiting fellowship during his sabbatical stay. IRS acknowledges support from the Australian Research Council Future-Fellowship Grant FT160100028.  CHIANTI is a collaborative project involving George Mason University, the University of Michigan (USA), University of Cambridge (UK) and NASA Goddard Space Flight Center (USA).

\end{acknowledgments}

\pagebreak

\input{table1.tex}

\input{table2.tex}

\clearpage
\bibliographystyle{apj}
\bibliography{SNRbib2}

\clearpage

\end{document}

%% file: table1.tex
\begin{deluxetable}{cccrrr}
\tablewidth{0pt}
\tablenum{1}
\tablecaption{Observed and Modeled Key Optical\tablenotemark{a} Emission Line Fluxes for P1 in Puppis A}
\tablehead{
\colhead {Ion} &  
\colhead {$\lambda$} &  
\colhead {Radial Velocity } &  
\colhead {Scaled} &
\colhead {Model }  \\
\colhead {} &  
\colhead {(\AA)} &  
\colhead {(km s$^{-1}$)} &  
\colhead {Int.\tablenotemark{b}} &
\colhead {A5\tablenotemark{d}} 
}
\startdata
\,Blue Channel  &   &     &   &   \\
\hline
\,[Fe III]     & 4658        &  $-$965 &    32   &    \nodata\\
\,H~I     & H$\beta$       &   $-$990 &   100   &   100       \\
\,[O III]     & 5007       &  $-$1035   &    17   &   12      \\
\,[Fe II]     & 5158        &  $-$1000   &    23   &   \nodata    \\
\,[N I]     & 5198+5200        &  $-$1040   &    43   &   46    \\
\hline
\,Red Channel  &   &     &   &   &\\
\hline
\,[N II]     & 5755       &  $-$1040   &    35   &  26  \\
\,[He I]     & 5876         &  $-$1020  &    18   &   9  \\
\,[N II]     & 6548         & $-$1005  &    280   & 205    \\
\,H~I     & H$\alpha$         &  $-$1020  &    100   &   100  \\
\,[S II]\tablenotemark{c}  & 6716        &  $-$990  &    50   &  49  \\
\,[Ar III] & 7135         &  $-$960  &    1.4   &   4 \\
\,[Fe II]     & 7155     & $-$1000 &    6   &   \nodata      \\
\,[Ca II]     & 7291           & $-$1000  &    9   & 2  \\
\,[Fe~II] & 8616         &  $-$990  &    3   &   4  \\
\,[Ni II]     & 7378         &  $-$1000  &    4.8   &   \nodata  &     \\
\enddata
\tablenotetext{a}{Emission line surface brightnesses scaled from the WiFeS spectrum with our adopted extinction correction applied; see text.}
\tablenotetext{b}{Surface brightness scaled to I(H$\beta$) = 100 (=3.5$\times$10$^{-16}$ ergs cm$^{-2}$ s$^{-1}$ arcsec$^{-2}$) for blue channel and I(H$\alpha$) = 100 (=1.2$\times$10$^{-15}$ ergs cm$^{-2}$ s$^{-1}$ arcsec$^{-2}$)  for red channel.}
\tablenotetext{c}{The [S~II] doublet ratio $\lambda$6716/$\lambda$6731 is 1.07, and  the model ratio is 1.08.}
\tablenotetext{d}{Model A5: 80 \kms\, shock, 8 $\rm cm^{-3}$, 90\% pre-ionized.  }
\label{tab1}
\end{deluxetable}

%% file: table2.tex
\begin{deluxetable}{cccrr}
\tablewidth{0pt}
\tablenum{2}
\tablecaption{Observed and Modeled Key Optical\tablenotemark{a} Emission Line Fluxes for Inner Ring Crosscut in Puppis A}
\tablehead{
\colhead {Ion} &  
\colhead {$\lambda$} &  
\colhead {Radial Velocity } &  
\colhead {Scaled} &
\colhead {Model }   \\
\colhead {D} &  
\colhead {(\AA)} &  
\colhead {(km s$^{-1}$)} &  
\colhead {Int.\tablenotemark{b}} &
\colhead {O\tablenotemark{d}} 
}
\startdata
\,Blue Channel  &   &     &   &   \\
\hline
\,[O II]\tablenotemark{c}     & 3727   & $-$765  &   1504 &    1250  \\
\,[Ne III]      & 3869  &    $-$650 &  358 &    340      \\
\,[O III]     & 4363       &   $-$695  &    108   &   150    \\
\,H~I     & H$\beta$       &   $-$780 &   100   &   100   \\
\,[O III]     & 5007       &  $-$725   &     2845   &    2670      \\
\hline
\,Red Channel  &   &     &   &   \\
\hline
\,He I     & 5876         &  $-$910  &    34   &   27       \\
\,[O I]     & 6300         & $-$770 &    134   &  $<$1      \\
\,H~I     & H$\alpha$         &  $-$780  &    100   &   100      \\
\,[N II]     & 6548         & $-$770  &    67   &   49       \\
\,[S II]\tablenotemark{c}  & 6716        &  $-$745  &    58   &   49      \\
\,[O II] & 7320         & $-$805  &    32   &   30       \\
\enddata
\tablenotetext{a}{Emission line surface brightnesses scaled from the WiFeS spectrum with our adopted extinction correction applied; see text}
\tablenotetext{b}{Surface brightness scaled to I(H$\beta$) = 100 (= 1.23$\times$10$^{-16}$ ergs cm$^{-2}$ s$^{-1}$ arcsec$^{-2}$) for blue channel and I(H$\alpha$) = 100 (= 3.72E-16 ergs cm$^{-2}$ s$^{-1}$ arcsec$^{-2}$) for red channel}
\tablenotetext{c}{Measured F($\lambda$3729)/F($\lambda$3727) = 1.7.  Model value = 1.3.}
\tablenotetext{c}{Measured F($\lambda$6716)/F($\lambda$6731) = 1.2.  Model value = 1.2.}
\tablenotetext{d}{Model O: 80 \kms\, shock, 50\% pre-ionized.  See text for details. }
\label{tab2}
\end{deluxetable}